# Design and Modeling of a Versatile Micro/Nanomotor Propulsion System by Light-Guided Dielectrophoresis

Zexi Liang, Donglei (Emma) Fan

*Abstract*—**To develop active materials that can efficiently respond to external stimuli with designed mechanical motions is one of the major obstacles that have hindered the realization of nanomachines and nanorobots. Here, we propose an innovative working mechanism that allows multifold-translational-motion control of semiconductor micro/nanomotors by AC dielectrophoresis with simple visible-light stimulation. We study the dielectrophoresis forces on semiconducting particles of various geometries in aqueous suspension by modeling with the consideration of both the Maxwell-Wagner relaxation and electrical-double-layer-charging effect. With the obtained understanding, we rationally design a manipulation system that can versatilely transport semiconductor micro/nanomotors and orient them towards desired directions at the same time by tuning the light intensity in an electric field. This research may guide the development of a new type of micro/nanomachine platform with high versatility and control. It is relevant to nanorobotics and nanodevice assembly.**

**INTRODUCTION**

The future micro/nanorobots require a high degree of freedom in motion control to perform complex tasks by individuals or in a swarm. However, on a miniaturized scale, it is a daunting task to apply established actuation approaches used in macroscale robots to micro/nanorobots. Once the dimension of a system is reduced to micro/nanoscale, the complexity in integrating multiple functional components into one structure, resembling those in bulk machines, increases dramatically. However, the level of required forces to drive a micro/nanorobot is much reduced. The interactions that are insufficient for macro robots can become prominent for micro/nanorobots. Such forces include those generated by external magnetic(*1-7*), electrical(*8, 9*), optical(*10*), and acoustic fields(*11*), as well as biochemical reactions(*12, 13*)), which have successfully compelled versatile locomotion of micro/nanomotors. However, it has been a daunting task to control the manipulation of individual micro/nanomotors within a swarm in the same manipulation platform, not to mention to achieve it agitely and reconfigurably with desired orientations(*14*).

Recently, we reported original working mechanisms that can rapidly switch the motion style and finely modulate the speed of both the mechanical rotation and alignment of a semiconductor micromotor in an AC electric field with simple visible light(*15, 16*). With experimentation and theoretical simulation, we clarified that it is the visible-light-induced photoconductivity that modulates both the real and imaginary parts of the electric polarizability of the silicon nanomotor at selected AC frequencies, which governs the electrorotation and electroalignment, respectively. The reconfigurable manipulations of the rotation and alignment of a nanomotor are demonstrated with many potential implementations, including the differentiation of metallic/semiconductor nanoparticles(*15*) and communication of information via Morse-coded mechanical motions(*16*). Nevertheless, how to realize the versatile translocation of individual nanomotors on a 2D surface in a swarm remains unknown and challenging.

Previously, the combined electrophoresis (EP) and dielectrophoresis (DEP) have been exploited to compel longitudinal micro/nanoparticles into linear transport with simultaneously controlled alignment(*17*). In a DC electric field, electrophoresis is originated from the coulomb and electrokinetic interactions between a surface-charged nanoparticle and an external DC field, transports the particle in the parallel or antiparallel direction of the DC field depending on the sign of its surface charge. In a high-frequency non-uniform AC electric field, a nanoparticle can experience DEP forces towards or away from the direction of the field gradient. The DEP force is governed by the real part of the electric polarizability of the nanoparticle. However, in a uniform AC electric field, the DEP force becomes zero, but still, the interaction can generate torque to align a particle independently. Thus, with a combined AC and DC electric field, a long particle can be manipulated on a 2D surface with transport and alignment controlled by the DC and AC fields, respectively. While, in this system, all particles behave the same, they synchronously move and align in the same manner.

In this work, we carry out modeling and simulation to study how light can be utilized to modulate the DEP force applied on silicon micro/nanoparticles of various geometries to realize versatile and individually controlled transportation. We unveil the complex working principle that depends on photoconductivity, AC frequency, as well as the shape and orientation of a particle. With the understanding, a new light-guided manipulation system is proposed that can control the transport direction, speed, and alignment of a particle all at the same time. (Fig. 1a).

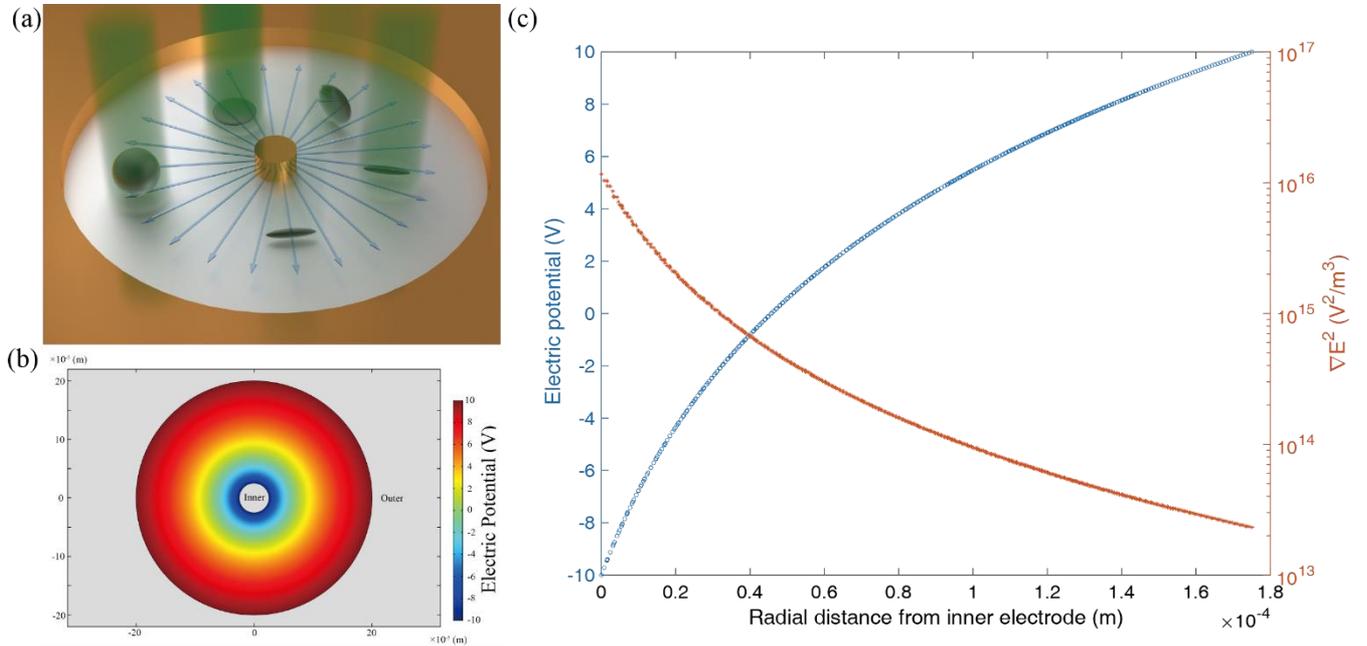

**Figure 1.** Schematic of light-guided dielectrophoretic propulsion of silicon nanoparticles of different geometries and numerical simulation of electric-field distribution. (a) A pair of concentric circular microelectrodes with inner and outer electrodes generates a non-uniform AC field (as indicated by the blue arrows). Silicon particles of various geometries in deionized (DI) water are propelled by the light-controlled DEP forces. Light patterns projected onto individual particles modulate their photoconductivities to control both the direction and magnitude of the DEP forces. (b) The mapping of electric potential distribution in the circular electrodes (two 10 Vpp sinusoudal signals applied with reversed phase.) (c) radial distribution of the electric potential and gradient of square of electric-field intensity ($\nabla_r E^2$) in a concentric circular electrode with an inner radius of 25 μm and an outer radius of 200 μm.

**Design of concentric microelectrodes to generate 1D radial electric-field and its gradient**

We design concentric circular microelectrodes to generate the required non-uniform electric field for the DEP force. This type of circular electrodes can be utilized for 1D manipulation along the radial direction (Figure 1a). The field can be generated via applying two AC signals with 180° phase shift to the inner (radius: 25 μm) and outer electrodes (radius: 200 μm), respectively. For instance, the voltage can be 10 Vpp. The electrical potential distribution when the phase of the AC voltage is zero is calculated as shown in Fig. 1b, c and the gradient of the square of electric-field intensity ($\nabla_r E^2$) is shown in Fig. 1c. This type of electrode can be readily utilized to verify the proposed working principle and demonstrate 1D manipulation.

**Fundamentals of DEP using the effective moment method**

When a nanoparticle is subjected to an electric field, an electrical polarization will be induced. The dipole moment is given by $\boldsymbol{p} = \alpha \boldsymbol{E}$, where α is the polarizability tensor of the particle. If the field is slightly non-uniform where a gradient exists, a force termed as dielectrophoretic force will exert on the particle, given by $\boldsymbol{F}_{DEP}(t) = \boldsymbol{p}(t) \cdot \nabla \boldsymbol{E}(t)$. For an AC electric field with a general expression of $\boldsymbol{E}(\boldsymbol{r},t) = \boldsymbol{E}_0(\boldsymbol{r},t) \cdot \exp(i\omega t)$, the DEP force is calculated as:(18)

$$\boldsymbol{F}_{DEP}(t) = \text{Re}\left[\underline{\boldsymbol{p}}\exp(i\omega t)\right] \cdot \nabla \text{Re}[\underline{\boldsymbol{E}_0}\exp(i\omega t)], \quad (1)$$

where $\underline{\boldsymbol{p}}$ and $\underline{\boldsymbol{E}_0}$ are the phasor of the dipole moment and electric field, respectively. Since the AC electric field period is generally much shorter than the time scale of the particle movement, what can be observed is the time-

averaged force ($< \boldsymbol{F}_{DEP}(t) >$) exerted on a particle as follows:

$$< \boldsymbol{F}_{DEP}(t) > = \frac{1}{2}\text{Re}\left[\boldsymbol{p} \cdot \nabla \boldsymbol{E_0^*}\right] = \frac{1}{2}\text{Re}(\underline{\alpha})\nabla E_{\text{rms}}^2, \qquad (2)$$

where the asterisk refers to complex conjugation, $\underline{\alpha}$ is the complex polarizability tensor of the particle, and $E_{\text{rms}}$ is the root mean square of the electric field intensity. A positive DEP force, which requires a positive real part of polarizability along the direction of the electrical field gradient, drags the particle towards the high-electric-field region and vice versa. If the particle is non-spherical with anisotropy in the electric polarizability, an alignment torque will be exerted on the particles, which is termed as electro-alignment. For example, if a particle changes its orientation within the x-y plane, the alignment torque can be given by:

$$\tau_{alignment} = -\frac{1}{2}E_0^2 \text{Re}[\alpha_x - \alpha_y]\sin\theta\cos\theta\ \hat{\boldsymbol{z}}, \qquad (3)$$

where $\theta$ is the angle between the electric field and the x-axis. When $\text{Re}(\alpha_x) > \text{Re}(\alpha_y)$, the particle's x-axis will orient towards the direction of the electric field, otherwise, the particle's y-axis will be aligned in the direction of the electric field.

Here, we specifically study ellipsoid particles with a, b, c as half the length of the principal axes for two reasons: 1) an analytical solution exists for ellipsoid particles when it is subjected to a uniform electric field along the principal axes. 2) a model of ellipsoid particles can predict the behavior of micro/nanoparticles of interest with shapes ranging from spheres, disks to rods, which are often made and used. In terms of the boundary value problem of an ellipsoid particle suspended in a uniform medium and a uniform electric field, an analytical solution exists. Suppose that the a, b, c axes of the ellipsoid particle are aligned along x, y, z axes of the coordinate system, respectively, the x component of the particle's electric polarization for example, can be expressed as:

$$p_x = \frac{4\pi abc}{3}\varepsilon_1 \frac{\underline{\varepsilon_2} - \underline{\varepsilon_1}}{\underline{\varepsilon_1} + (\underline{\varepsilon_2} - \underline{\varepsilon_1})L_x}E_x = \alpha_x E_x, \qquad (4)$$

where $\varepsilon_1$ is the real permittivity of the medium, $\underline{\varepsilon_1}, \underline{\varepsilon_2}$ are the complex permittivity of the medium and the particle, respectively, given by $\underline{\varepsilon_i} = \varepsilon_i - \frac{i\sigma}{\omega}$, and $L_x$ is the depolarization factor defined by the following elliptical integral: $L_x = \frac{abc}{2}\int_0^\infty \frac{ds}{(s+a^2)\sqrt{(s+a^2)(s+b^2)(s+c^2)}}$.

Similarly, the expression for $p_y$ and $p_z$ can be obtained by the substitution of $L_x$ and $E_x$ in Equation (4) with the corresponding items in the respective direction.

**Frequency dependent electric polarization**

The frequency-dependent electric polarization of dielectric materials is known as the dispersion. In optical frequencies, the Lorentz model can well explain the dispersion relation of dielectric materials originate from multiple relaxation mechanisms. In most dielectrophoresis experiments, however, the frequency range of the alternating electric field is much lower and falls in the range of kHz to MHz, where the relaxation processes of atomic dipole, ionic polarization and electronic polarization are not observed(*19*). The only possible relaxation mechanism in these frequencies is the interface relaxation, also known as the Maxwell-Wagner interfacial polarization, which is originated from the discontinuity of the electrical properties at the interface between two mediums. Besides the effect of Maxwell-Wagner polarization that has been calculated in equation (4), due to the existence of ions in aqueous solution, once the nanowire is polarized, the ions of opposite charges are spontaneously attracted towards the polarized surface to counter the surface charge and thus form charge electric layers, the so-called electrical double layers. The behavior of the electrical double layer in an AC electric field could be very complex and several models have been developed in order to predict the electrokinetic phenomena.(*20-22*) Here we simplify the system with an RC model for qualitative analysis only and the dipole moment contributed by the electrical double layer, expressed as(*15*):

$$\text{Im}(P_{EDL}) = -\frac{[\text{Re}(P_{MW})\sin\delta + \text{Im}(P_{MW})\cos\delta]}{\sqrt{\omega^2\tau_{RC}^2 + 1}} \qquad (5)$$

$$\text{Re}(P_{EDL}) = -\frac{[\text{Re}(P_{MW})\cos\delta - \text{Im}(P_{MW})\sin\delta]}{\sqrt{\omega^2\tau_{RC}^2 + 1}} \qquad (6)$$

where $\tau_{RC}$ is the time constant for EDL charging and $\tan\delta = -\omega\tau_{RC}$ as the dielectric loss tangent.

Both the Maxwell-Wagner polarization and the electrical double layer contribute to the total electric dipole moment of the particles. Both the real and imaginary parts of the total dipole moment can be calculated accordingly. In this study, we select silicon as the material of the particle and deionized water as the suspension medium. Silicon is chosen due to the earthly abundance, high photoconductivity, and relevant to various practical applications. To understand the effect of light on the modulation of the DEP forces on Si nanoparticles via the tuning of their electric conductivities, we systematically sweep the electrical conductivity from 0.001 S/m to 1 S/m in the calculation.

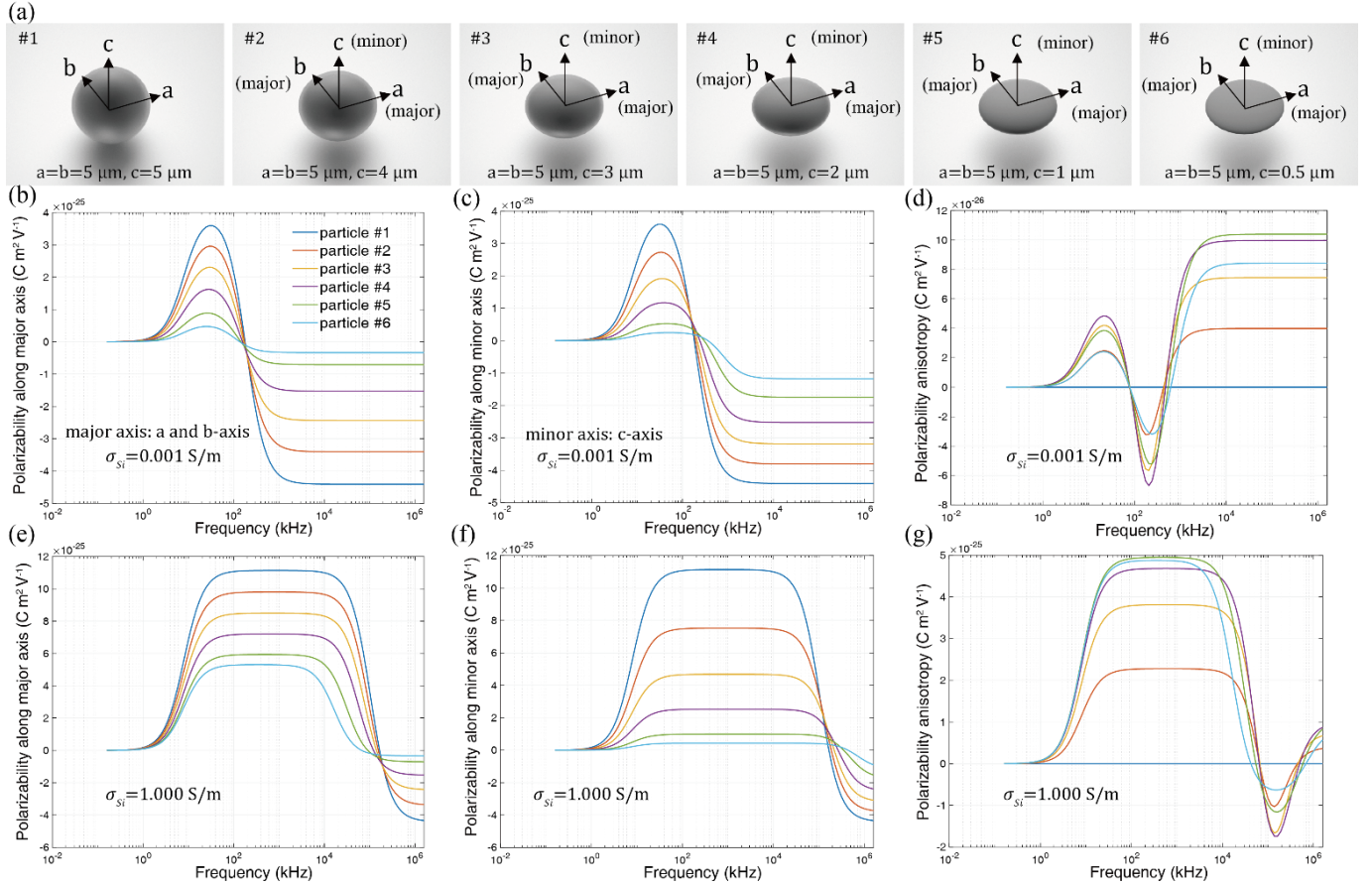

**Figure 2.** Calculation of the electric polarizability and anisotropy (difference between polarizabilities along major axis and minor axis) of various oblate spheroid Si particles with different electric conductivities. (a) Six oblate spheroid particles used in the calculation. (b-d) The electric polarizability along the major axes (a and b-axis), minor axis (c-axis) and the anisotropy of the oblate spheroid particles with a low conductivity of 0.001 S/m, and (e-g) a high conductivity of 1 S/m.

## Calculation and analysis

We first investigate the geometry dependence of the electric polarization. A series of oblate spheroid silicon particles ($a = b > c$) are calcualted, including one sphere ($a = b = c = 5\ \mu m$) and another 5 spheroids with monotonically reduced z dimension from $c = 4\ \mu m$ to $c = 0.5\ \mu m$ ($a = b = 5\ \mu m$) at a low ($\sigma_{Si} = 0.001$ S/m) and high ($\sigma_{Si} = 1$ S/m) electric conductivity, respectively. The polarizability is calculated by using the model discussed above by considering both Maxwell-Wagner relaxation and the electrical-double-layer effect. As shown in Fig. 2b-c, the electric polarizability along the major axes (the longest principal axes, which are a-axis and b-axis in this case) always vanishes at the low frequency limit due to the EDL screening effect, while at the high-frequency range, it levels off to a negative value, which results from the Maxwell-Wagner relaxation at the high frequency limit, since the EDL effect gradually fades with the increase of AC frequency. The magnitude of the electric polarizability decreases when the spheroid shrinks in the c-axis, which agrees with equation (3), which indicates the electric polarization is directly proportional to the volume of a particle. By comparing the electric polarizability at both low and high electric conductivities controlled by light illumination, it can be readily found that the electric conductivity plays an imperative role in governing the frequency-dependent electric polarization. First of all, by increasing the electric conductivity, the magnitude of the peak of the positive electric polarizability can be augmented by several folds depending on the particle geometry, and the peak of the positive polarization can extend to a broad range of frequencies into a plateau. By tuning the electric conductivity from 0.001 S/m to 1 S/m, for the spherical particle

(#1), the maximum positive polarizability can be magnified by around 3 times, while for the disk-like oblate spheroid (#6), the magnification is more than 11 times (Table 1). Thus, with the reduction of the dimension along the c axis, the increase of the electric polarization by the light-enhanced electric conductivity becomes more effective. For the negative polarizability, at the low electric conductivity, the switching of the electric polarizability from positive to negative occurs at around 200 kHz, while at the high electric conductivity controlled by light, the switching frequency shifts to above 100 MHz, which is already beyond the commonly used AC frequency range for electrokinetics. As a result, below 200 kHz, the positive electric polarization can be much improved by the increase of the electric conductivity, and the positive DEP force can be amplified several times. At the same condition, when it is above 200 kHz, the sign of the electric polarization, as well as the DEP force, will be switched from negative to positive.

Furthermore, when a particle changes the geometry from spherical, oblate spherical, to disk as shown in Fig. 2(a), the directional anisotropy of electric polarizability emerges, and the electric polarizability along the c-axis differs from that of a-axis and for a particle of either a low (Fig. 2d) or a high electric conductivity (Fig. 2g). The electric polarization along the minor axis is largely similar to that along the major axis for most oblate spheroids, however, a difference is that the light-induced magnification of the positive electric polarization along the minor decreases with the decrease of the dimension in the minor axis, opposite to the electric polarizability along the major axis. This indicates that for an oblate spheroid, the polarization along the major axes (a-axis and b-axis) is more sensitive to the change of electric conductivity compared to that along the minor axis (c-axis). As the anisotropy of the electric polarizability along different axes emerges, the orientation of the particle should be considered since the total electrical potential energy will be different along different orientations. For a oblate spheroid, it will either orient with a,b-axis or c-axis along the direction of the electric field depending on which polarizability has a greater real-part value as shown in equation (3). We calculate the anisotropy of the electric polarizability, given by $\text{Re}[\alpha_a - \alpha_c]$, of all the six shapes (Fig. 2a) for a particle of either a low (Fig. 2d) or a high electric conductivity (Fig. g). It is found that at a low electric conductivity, all the oblate spheroidal particles orient towards the a-axis (or b-axis) along the electric-field direction below ~77 kHz, and switch to the c-axis alignment with the electric field between 77 kHz to around 550 kHz (slightly different for different geometries), and then switch back to the a-axis alignment at even higher frequencies. At a high conductivity, all of the oblate spheroidal particles align with one of its major axes orienting along the electric-field direction below 60 MHz, and switch to the minor-axis-alignment in the range of ~ 60 MHz to ~100 MHz, and then switch back to the major-axis-alignment at even higher frequencies. The sphere (#1 in Fig. 2(a)) does not show electric anisotropy and alignment effect, but even a light shape change can result such an effect (#2-6 in Fig. 2(a)). In addition, by comparing the sign of the anisotropy of the electric polarizability between 77 kHz to 550 kHz for a particle of a low (Fig. 2d) and high conductivity (Fig. 2g), it is clear that the orientation of oblate particles can be facilely switched between the two alignment states by tuning the electric conductivity with light. More will be discussed later.

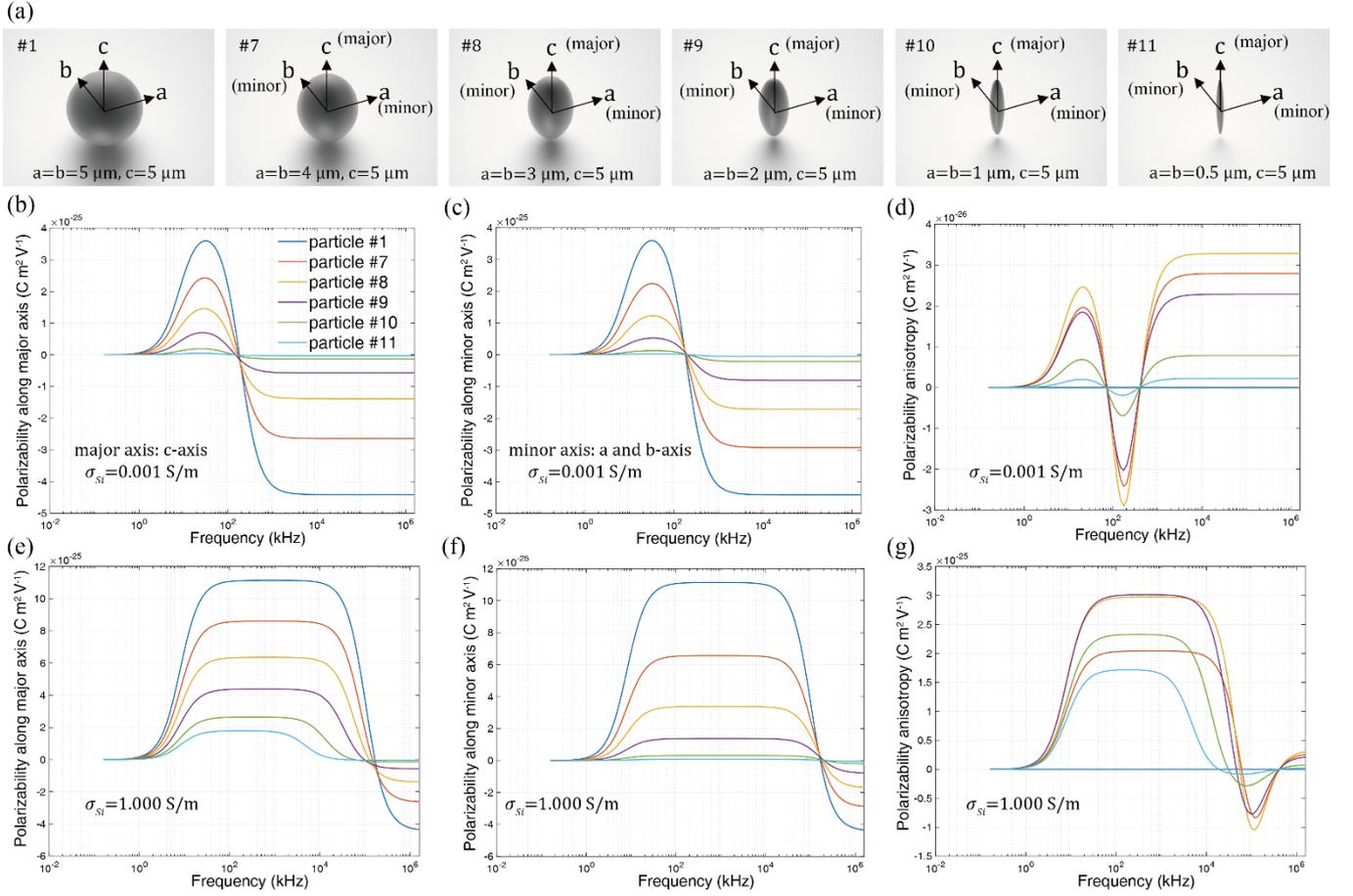

**Figure 3.** Calculation of electric polarizability and electric anisotropy of various prolate spheroidal silicon particles with different electric conductivities. (a) Diagrams of six prolate spheroidal particles. (b-d) Electric polarizability along the major axes (c-axis) and minor axes (a and b-axis), and the electric anisotropy of the six prolate spheroid particles of low electric conductivity of 0.001 S/m and (e-g) high conductivity of 1 S/m.

Another type of special geometry of interest is the prolate spheroids (a=b<c). Here we studied 6 prolate spheroids with the half length $a$ and $b$ ranging from 5 μm to 0.5 μm, and $c = 5$ μm. As shown in Fig. 3a-f, the prolate spheroid polarizability spectrum is generally similar to that of the oblate spheroid, with moderate differences in amplitude and frequency. Particularly, it is found that the amplification of the positive electric polarizability along the major axis (c-axis) between the low and high conductivities increases with the aspect ratio $\frac{c}{a}$. Particle #11 for example, has an aspect ratio of 10, and its polarizability along the c-axis amplifies more than 35 times as the electric conductivity increases from 0.001 to 1 S/m by light, while with the same change of electric conductivity, the electric polarizability of a spherical particle (#1) only can be enhanced by around 3 times (Fig. 3b and e, Table 1). Thus, the prolate spheroids are better shapes for obtaining a large range of light-tunable DEP forces with high sensitivity to electric conductivity. However, in terms of switching the moving direction of the prolate spheroids, it can be seen that a more spherical-shaped particle exhibits closer values in the positive and negative electric polarizabilities with and without light, respectively, if aligned along the c-axis, e.g. at around 5x10³ Hz. This type of geometry could be helpful for obtaining flexibly controlled 2D transport.

| Particle number (#) | 1 | 2 | 3 | 4 | 5 | 6 |
|---|---|---|---|---|---|---|
| Maximum positive polarizability at low conductivity ($Cm^2V^{-1}$) | $3.59 \times 10^{-25}$ | $2.96 \times 10^{-25}$ | $2.31 \times 10^{-25}$ | $1.63 \times 10^{-25}$ | $8.84 \times 10^{-26}$ | $4.69 \times 10^{-26}$ |
| Maximum positive polarizability at high conductivity ($Cm^2V^{-1}$) | $1.11 \times 10^{-24}$ | $9.81 \times 10^{-25}$ | $8.49 \times 10^{-25}$ | $7.20 \times 10^{-25}$ | $5.94 \times 10^{-25}$ | $5.31 \times 10^{-25}$ |
| Magnification | 3.1x | 3.3x | 3.7x | 4.4x | 6.7x | 11.3x |

| Particle number (#) | 7 | 8 | 9 | 10 | 11 |
|---|---|---|---|---|---|
| Maximum positive polarizability at low conductivity ($Cm^2V^{-1}$) | $2.43 \times 10^{-25}$ | $1.46 \times 10^{-25}$ | $6.99 \times 10^{-26}$ | $1.92 \times 10^{-26}$ | $5.01 \times 10^{-27.}$ |
| Maximum positive polarizability at high conductivity ($Cm^2V^{-1}$) | $8.60 \times 10^{-25}$ | $6.36 \times 10^{-25}$ | $4.39 \times 10^{-25}$ | $2.64 \times 10^{-25}$ | $1.79 \times 10^{-25}$ |
| Magnification | 3.5x | 4.4x | 6.6x | 13.8x | 35.7x |

Table 1. The maximum positive polarizability of all particles along the major axis at low and high conductivities and the magnification factors.

The above discussion summarizes the systematic study of the dependence of electric polarizabilities of silicon particles on geometry. The obtained understanding is essential to realize the function-oriented design of micro/nanoparticles for efficient light-reconfigurable manipulation and application.

**Light-stimulated versatile transport of microparticles with simultaneously controlled alignment**

The results of the dependence of the electric polarizability of Si microparticles on the geometry, AC frequency and electric conductivity are critical for us to design innovative systems for light-stimulated versatile transport of microparticles with simultaneously controlled alignment. For simplicity, we first consider the scenario that the electric-field gradient is parallel to the electric field direction. Such an electric-field distribution can be implemented by applying an AC voltage onto two concentric circular electrodes as shown in Fig. 1a, b. Both the electric field and the gradient of the electric field are radial, pointing towards the center of the inner electrode (to be quantitatively discussed later). We select particle #1 (sphere), #6 (oblate spheroid, disk-like) and #11 (prolate spheroid, rod-like) as three representative geometries and demonstrate versatile manipulation in both transport and alignment by the frequency and photoconductivity modulation.

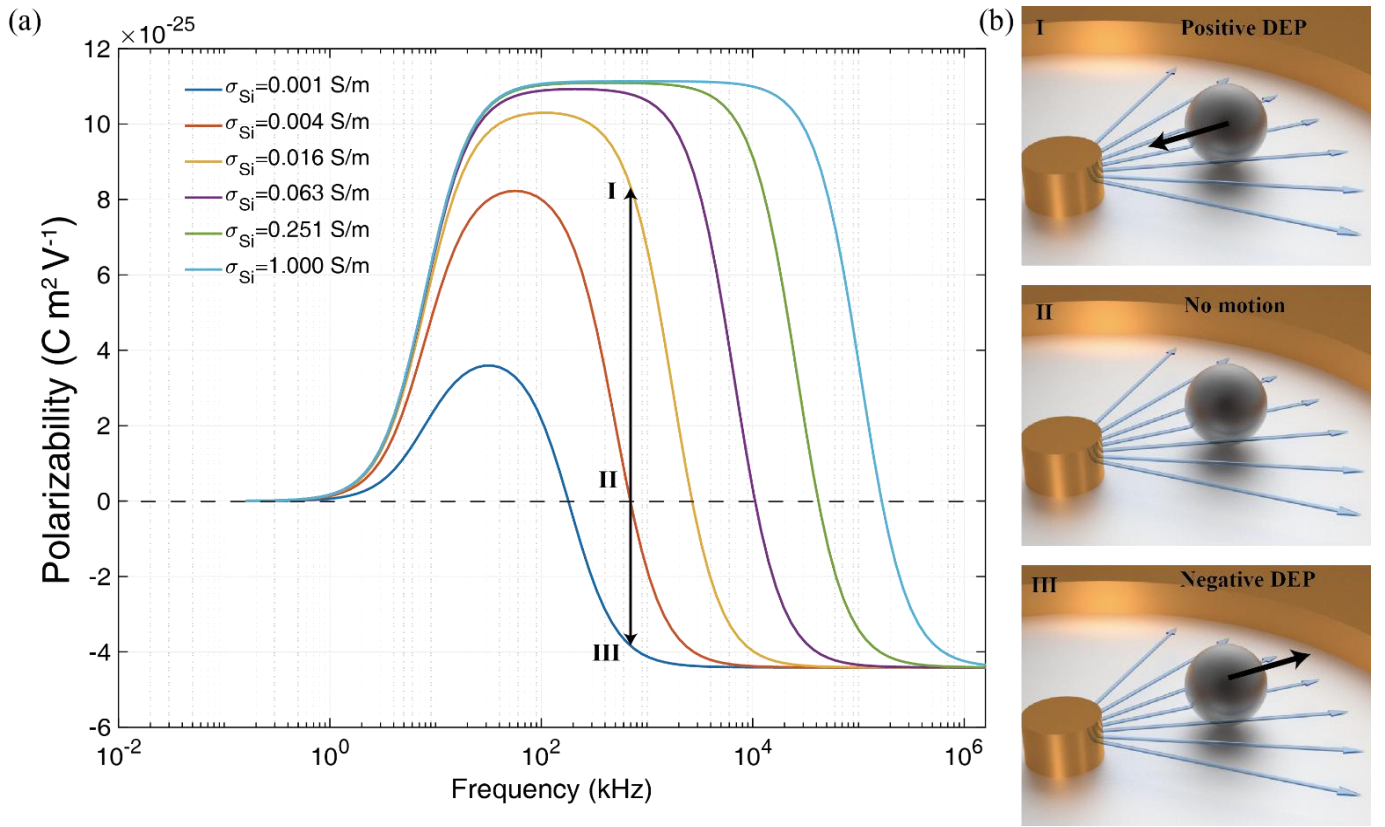

**Figure 4.** Calculation of the electric polarizability of the spherical particle (#1). (a) Electric polarizability of a spherical particle versus AC frequency with different electric conductivities. At 700kHz, I, II and III show three possible motion modes with positive, zero and negative DEP forces, respectively. (b) Corresponding diagrams of the three motion modes.

Due to the geometric symmetry, a spherical particle experiences zero alignment torque in an AC electric field, and the polarizability is isotropic. Within a concentric electric field, there only are three possible modes of motion for the spherical particles: (1) moving towards the center, (2) moving away from the center (3) no motion. It is possible to switch the motion modes by just controlling the light intensity at a given AC frequency, a result of the tuning of the photoconductivity of silicon particles. Here, in addition to the light-controlled photoconductivity, there could be additional electric conductivity from doping and thermal excitation. For simplicity, we consider the Si is undoped and the photoconductivity is the dominant source of electric conductivity due to its many orders of magnitude change. We calculate the electric polarizability of the spherical particle #1 by varying the electric conductivities from 0.001 S to 1 S as shown in Fig. 4a. The AC frequency is selected to be 700 kHz. A base light intensity is defined at which the particle's electric conductivity equals 0.004 S/m. At this condition, the real-part electric polarizability equals zero and the particle remains still as indicated as the point II in Fig.4a and Fig. 4c. If the light intensity increases higher than this base value, a positive DEP force will be triggered, and the particle will transport inwards to the center electrode (point I in Fig. 4a and Fig. 4b). If the light intensity decreases below this base value, a negative DEP force will be generated, and the particle will move outwards (point I in Fig. 4a and Fig. 4d).

For the oblate particle #6, there are more possible modes of motion. To depict the general behavior, we first calculate the real-part polarizability along the major axes (a,b-axis) and minor axis (c-axis) as well as the polarizability anisotropy ($\text{Re}[\alpha_a - \alpha_c]$) with a series of electric conductivities from 1 S/m to 0.001S/m as shown in Fig. 5a-c. With the obtained results, we select an AC frequency at 1 MHz, where we can obtain five distinct motion modes as summarized in Fig. 5d and Table 2.

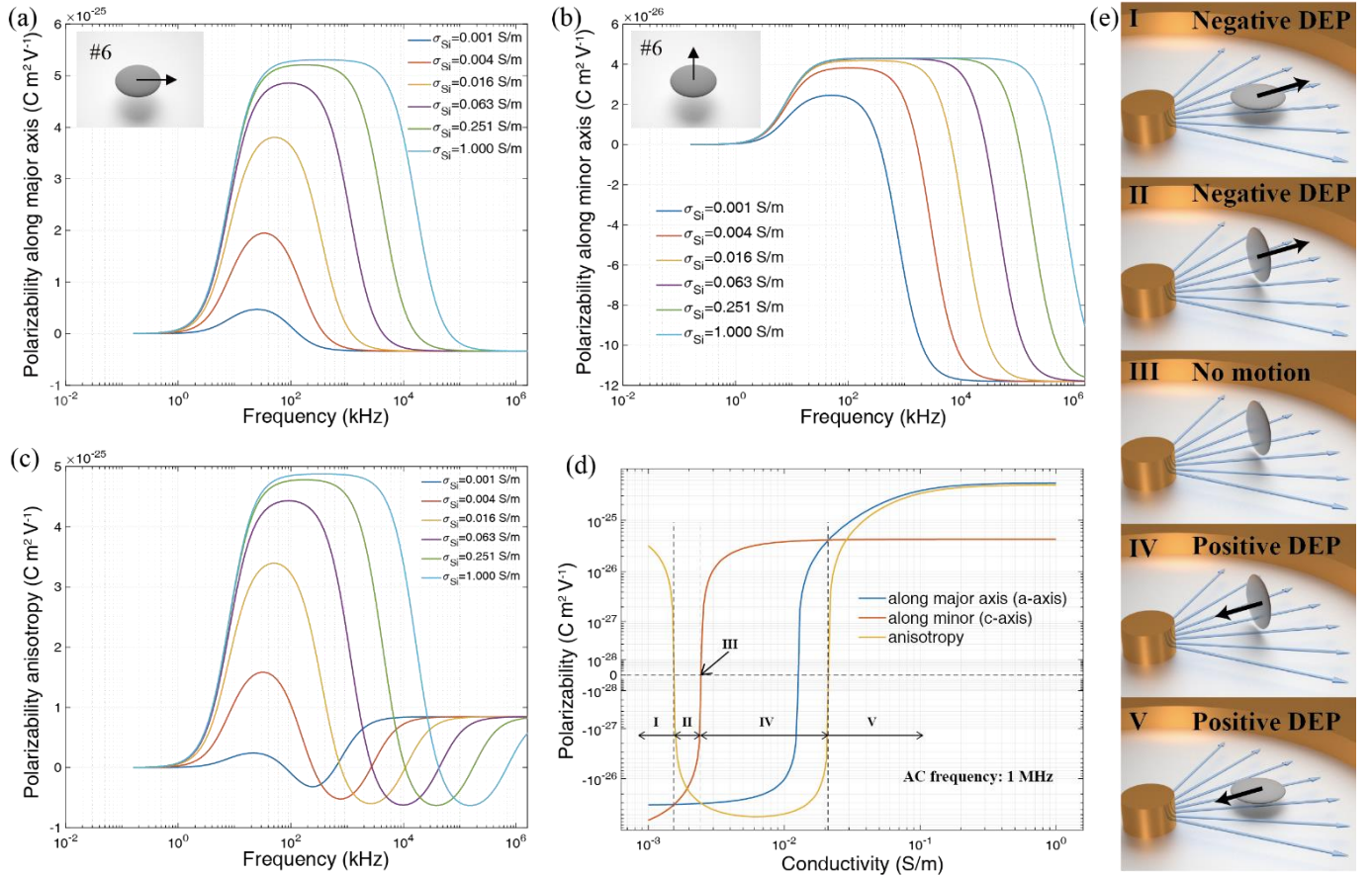

**Figure 5.** Calculation of the electric polarizability of the oblate spheroid particle (#6). Electric polarizability of the particle along (a) the major axis and (b) the minor axis versus AC frequency with different electric conductivities. (inset: diagram of the polarization direction). (c) Electric polarizability anisotropy between the major-axis and minor-axis polarizabilities. (d) Electric polarizability and anisotropy as a function of electric conductivity at 1 MHz. (e) Diagrams of the five inter-switchable motion modes.

| Motion Modes | I | II | III | IV | V |
|---|---|---|---|---|---|
| Conductivity (S/m) | <0.0015 | 0.0015-0.0024 | ~0.0024 | 0.0024-0.0219 | >0.0219 |
| Orientation along E-field | major-axis | minor-axis | minor-axis | minor-axis | major-axis |
| Sign of DEP force | negative (-) | negative (-) | zero | positive (+) | Positive (+) |

Tabel 2. Five inter-switchable motion modes of the oblate silicon particle (#6) with tunable electric conductivity by light in an electric field at 1 MHz

With the same method, we analyze the electric polarizability, anisotropy, and motion modes of the prolate particle #11 as shown in Fig. 6a-d in Table 3. For the prolate particle #11, a, b-axis are the minor axes, and c-axis is the major axis. The anisotropy is defined as $\text{Re}[\alpha_c - \alpha_a]$. The selected AC frequency is 2 MHz.

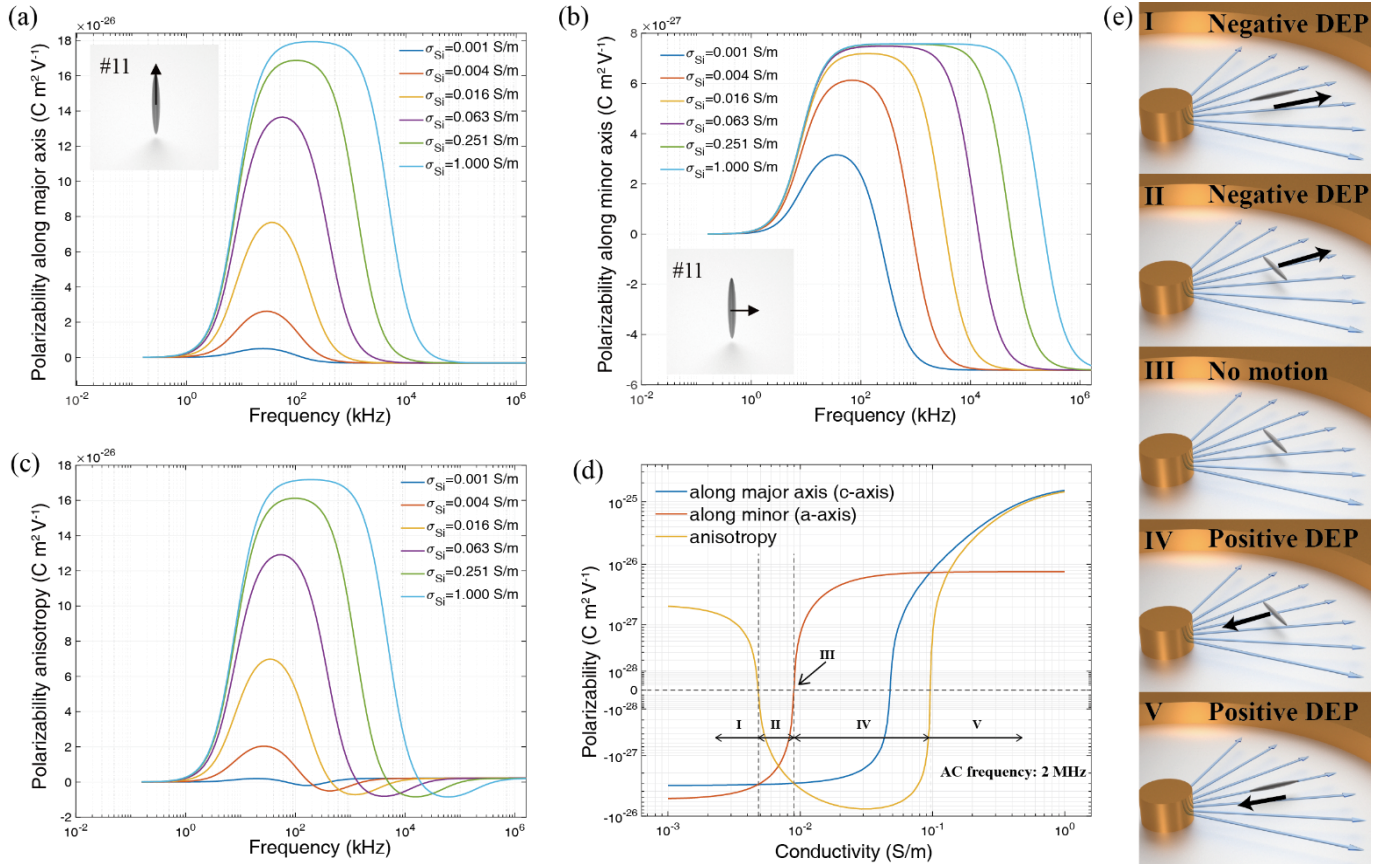

Figure 6. Calculation of the electric polarizability of the prolate spheroid particle (#11). Electric polarizability of the particle along (a) the major axis and (b) the minor axis versus AC frequency with different electric conductivities. (inset: diagram of the polarization direction). (c) Electric polarizability anisotropy between the major-axis and minor-axis polarizabilities. (d) Electric polarizability and anisotropy as a function of electric conductivity at 2 MHz. (e) Diagrams of the five inter-switchable motion modes.

| Motion modes | I | II | III | IV | V |
|---|---|---|---|---|---|
| Conductivity (S/m) | <0.0024 | 0.0024-0.0045 | ~0.0045 | 0.0045-0.0437 | >0.0437 |
| Orientation along E-field | major-axis | minor-axis | minor-axis | minor-axis | major-axis |
| **Sign of DEP force** | negative | negative | zero | positive | positive |

Tabel 3. Five inter-switchable motion modes of the oblate silicon particle (#11) with tunable electric conductivity by light in an electric field at 2 MHz.

In summary, we demonstrate that for both the prolate and oblate spheroid particles, by selecting proper AC frequencies, one can rationally select and switch the translocation direction and alignment of a particle among 5 possible modes by simply modulating the photoconductivity. Two orientations, either alignment along the semi-major or the semi-minor axis with the electric field can be obtained. Within each orientation, the DEP force can be finely tuned from a negative value to a positive value.

**Discussion and Conclusion**

In this work, we propose a new and versatile working mechanism that could be applied for developing an array of semiconductor-based light-guided dielectrophoresis manipulation system. The electrical property of a semiconductor micromotor can be controlled by external light stimuli and is exhibited as tunable mechanical motions in an external AC electric field. Systematic investigation of the polarization behavior of semiconductor particles has been conducted, focusing on the dependence on the geometry, AC frequency, and electric conductivity. The selection and switching of modes of mechanical motion with only light modulated electric conductivity has been demonstrated on spherical, oblate and prolate particles. Although the manipulation is 1D transport in circular microelectrodes, the same working principle could be extended to 2D and 3D manipulations by utilizing strategically designed microelectrodes to generate required field gradient in more dimensions. This work could be particularly interesting for applications related to the translocation of micro/nanomotors (or robots), such as the delivery of cargos to live cells in biological study.  It also could be utilized as a general platform that coordinates activities of nanomotor swarms for a collaborative task force. Furthermore, it may be exploited for the rapid assembly of artificial particles with dually controlled alignment and position for various electronic applications.

**Acknowledgment**

We are grateful for the support of NSF via the CAREER Award (grant no. CMMI 1150767 and intern supplement) and research grants (1710922 and 1930649), the support of the Welch Foundation (grant no. F-1734) and the University Graduate Continuing Fellowships to ZL (The University of Texas at Austin, 2019).

**References**

1. Z. Wu et al., A swarm of slippery micropropellers penetrates the vitreous body of the eye. *Science Advances* **4**, eaat4388 (2018).
2. A. Ghosh, P. Fischer, Controlled Propulsion of Artificial Magnetic Nanostructured Propellers. *Nano Letters* **9**, 2243-2245 (2009).
3. J. Giltinan, M. Sitti, Simultaneous Six-Degree-of-Freedom Control of a Single-Body Magnetic Microrobot. *IEEE Robotics and Automation Letters* **4**, 508-514 (2019).
4. S. Tottori et al., Magnetic helical micromachines: fabrication, controlled swimming, and cargo transport. *Advanced materials* **24**, 811-816 (2012).
5. B. Wang et al., Reconfigurable Swarms of Ferromagnetic Colloids for Enhanced Local Hyperthermia. *Advanced Functional Materials*, 1705701 (2018).
6. J. Cui et al., Nanomagnetic encoding of shape-morphing micromachines. *Nature* **575**, 164-168 (2019).
7. T. Li et al., Highly efficient freestyle magnetic nanoswimmer. *Nano letters* **17**, 5092-5098 (2017).
8. K. Kim, X. Xu, J. Guo, D. Fan, Ultrahigh-speed rotating nanoelectromechanical system devices assembled from nanoscale building blocks. *Nature communications* **5**, 3632 (2014).
9. U. Ohiri et al., Reconfigurable engineered motile semiconductor microparticles. *Nature Communications* **9**, 1791 (2018).
10. L. Wang, Q. Li, Photochromism into nanosystems: towards lighting up the future nanoworld. *Chemical Society Reviews* **47**, 1044-1097 (2018).

11. L. Ren, W. Wang, T. E. Mallouk, Two forces are better than one: combining chemical and acoustic propulsion for enhanced micromotor functionality. *Accounts of chemical research* **51**, 1948-1956 (2018).
12. L. Ren *et al.*, Rheotaxis of Bimetallic Micromotors Driven by Chemical–Acoustic Hybrid Power. *ACS nano* **11**, 10591-10598 (2017).
13. W. Wang *et al.*, A tale of two forces: Simultaneous chemical and acoustic propulsion of bimetallic micromotors. *Chem Commun* **51**, 1020-1023 (2015).
14. G.-Z. Yang, P. Fischer, B. Nelson, New materials for next-generation robots. *Science Robotics* **2**, eaap9294 (2017).
15. Z. Liang, D. Fan, Visible light–gated reconfigurable rotary actuation of electric nanomotors. *Science advances* **4**, eaau0981 (2018).
16. Z. Liang, D. Teal, D. E. Fan, Light programmable micro/nanomotors with optically tunable in-phase electric polarization. *Nature communications* **10**, 1-10 (2019).
17. D. Fan, R. Cammarata, C. Chien, Precision transport and assembling of nanowires in suspension by electric fields. *Applied Physics Letters* **92**, 093115 (2008).
18. T. B. Jones, T. B. Jones, *Electromechanics of particles*. (Cambridge University Press, 2005).
19. J. D. Jackson, *Classical electrodynamics*. (John Wiley & Sons, 2007).
20. M. Z. Bazant, T. M. Squires, Induced-charge electrokinetic phenomena: theory and microfluidic applications. *Physical Review Letters* **92**, 066101 (2004).
21. P. García-Sánchez, J. E. Flores-Mena, A. Ramos, Modeling the AC electrokinetic behavior of semiconducting spheres. *Micromachines* **10**, 100 (2019).
22. T. Miloh, Nonlinear alternating electric field dipolophoresis of spherical nanoparticles. *Physics of fluids* **21**, 072002 (2009).